\documentclass[twocolumn,aps,prl,showpacs,preprintnumbers,amsmath,amssymb, superscriptaddress]{revtex4}

\usepackage{graphicx}
\usepackage{dcolumn}
\usepackage{bm}

\begin{document}


\title{Localization in
one-dimensional incommensurate lattices beyond the Aubry-Andr\'e model }


\author{J. Biddle}
\affiliation{Condensed Matter Theory Center, Department of
Physics, University of Maryland, College Park, Maryland 20742, USA}
\author{B. Wang}
\affiliation{Condensed Matter Theory Center, Department of
Physics, University of Maryland, College Park, Maryland 20742, USA}
\author{D. J. Priour, Jr}
\affiliation{Condensed Matter Theory Center, Department of
Physics, University of Maryland, College Park, Maryland 20742, USA}
\affiliation{Physics Department, University of Missouri, Kansas City, Missouri 64110, USA}
\author{S. Das Sarma}
\affiliation{Condensed Matter Theory Center, Department of
Physics, University of Maryland, College Park, Maryland 20742, USA}


\date{\today}

\begin{abstract}
Localization properties of particles in one-dimensional
incommensurate lattices without interaction are
investigated with models beyond the tight-binding Aubry-Andr\'e
(AA) model. Based on a tight-binding $t_{1} - t_{2}$ model with
finite next-nearest-neighbor hopping $t_{2}$, we find the
localization properties qualitatively different from those of the
AA model, signaled by the appearance of mobility edges. We then
further go beyond the tight-binding assumption and directly study
the system based on the more fundamental single-particle
Schr\"odinger equation.  With this approach, we also observe the presence of mobility edges and localization properties dependent on incommensuration.  
\end{abstract}

\maketitle

    The physics of quantum transport in random disordered
    potentials has been a subject of substantial interest for
    condensed matter physicists for decades. The extended Bloch
    waves in a periodic lattice could undergo a quantum interference induced transition into
    localized states due to random disorder by a mechanism
    commonly referred to as Anderson localization \cite{Anderson}.    Matter waves can also be localized in deterministic potentials
    that exhibit some similarities to random disorder
    \cite{Aubry, DasSarma90, DasSarma88, Thouless88}. Quasi-periodic
    potentials, such as incommensurate lattices (the superposition
    of two or more lattices with incommensurate periods), are
    notable examples and have been extensively studied with the
    Aubry-Andr\'e model \cite{Aubry}. Such potentials have been
    shown to exhibit interesting quantum transport phenomena in
    themselves. Incommensurate potentials, for example, are
    theorized to have fractal spectrums \cite{Saul88}.    However, it remains challenging to study these phenomena in
    solid state experiments, as it is difficult to systematically
    control the disorder in solid state systems. In contrast to
    the solid state systems, ultracold atoms loaded in optical
    lattices offer remarkable controllability over the system
    parameters, making it an attractive platform for the study of
    the localization of matter waves. Recently, Anderson
    localization of noninteracting Bose-Einstein condensates (BEC)
    has been observed in a one-dimensional matter waveguide with a
    random potential introduced with laser speckles
    \cite{Billy08}. Similar experiments have also been done in
    quasi-periodic optical lattices\cite{Edwards08, Roati08}.

    Localization of noninteracting particles in one dimensional
    incommensurate lattices is often studied with the
    Aubry-Andr\'e model (AA) with nearest neighbor (nn) hopping,
    where one of the lattices is assumed to be relatively weak and
    can be treated as a perturbation . Within the framework of
    the AA model, there is a duality point, at which a sharp
    transition from all eigenstates being extended to all being
    localized occurs. However, in ultracold atom experiments, one
    can tune the depth of each lattice in a controllable way and
    bring the system out of the tight-binding regime. To explore
    the physics of localization for shallow lattices, it is of
    interest to go beyond the AA model and the tight-binding
    assumption \cite{Diener01}.

    In this work, we first study the tight-binding $t_{1} - t_{2}$
    model, which extends the AA model by including the
    next-nearest neighbor (nnn) hopping. The inclusion of the nnn hopping
    destroys the self-duality possessed by the AA model and the
    localization properties of the system become more complex through
    the emergence of mobility edges. We then examine the system
    directly with the single particle Schr\"odinger equation. We
    discretize the equation and solve it numerically without any
    further assumption. Within this formalism, we also find the
    existence of mobility edges, consistent with the $t_{1} -
    t_{2}$ model results, and we find localization properties with non-trival dependence on incommensuration.

    Consider diffuse, noninteracting, ultra-cold atoms in a
    one-dimensional incommensurate lattice, where the
    atoms can only move along the x-axis. The lattice potential
    is given by
\begin{equation}
V(x)=\frac{V_0}{2}{\rm \cos}(2k_Lx)+\frac{V_1}{2}{\rm
\cos}(2\alpha k_Lx+\delta), \label{eq:latpot}
\end{equation}
where $V_{0}$ and $V_{1}$ describe the depth of the primary and
secondary lattices respectively, $k_{L}$ is the wave-vector of the
primary lattice along the x-axis, $\alpha$ is an irrational number
characterizing the degree of incommensurability between the
periods of the two lattices, and $\delta$ is an arbitrary phase
(in our calculations it is chosen to be zero for convenience,
without loss of generality). When the depth of the primary lattice
is sufficiently large as compared with the recoil energy
$E_r\equiv(\hbar k_{L})^{2}/2m$ as well as the depth of the
secondary lattice $V_1$, the physical properties of the system can
be studied with the well-known single-band tight-binding
Aubry-Andr\'e model:
\begin{equation}
t(u_{n-1}+u_{n+1})+V_nu_n=Eu_n, \label{eq:AAmodel}
\end{equation}
in which only the coupling between nearest-neighbors (nn) is
retained and the incommensurate modulating potential $V_n=V\cos(2\pi\alpha n)$.
The duality point is given by $V/t = 2$. The nn hoping term, $t$,
is determined by the primary potential and can be approximated by
the expression
\begin{equation}
t\approx\frac{4}{\sqrt{\pi}}E_r(\frac{V_0}{E_r})^{3/4}{\rm
exp}(-2\sqrt{\frac{V_0}{E_r}}), \label{eq:teq}
\end{equation}
\begin{figure}
\includegraphics[width=.45\textwidth]{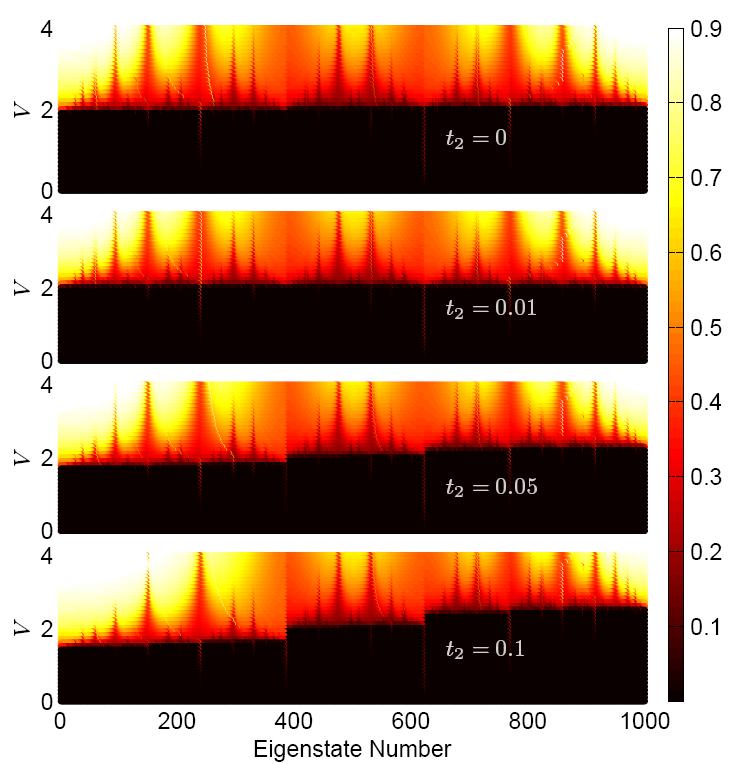}

\caption{\label{fig:Fig1} Inverse participation ratio of all
eigenstates for $t_1-t_2$ model with $\alpha =(\sqrt{5}-1)/2$.
The size of the system is chosen to have 1000 sites. The four panels
correspond to $t_2$ = 0, 0.01, 0.05, and 0.1 respectively. ($t_1$
is the unit for energy.) Darker shading corresponds to more extended states while lighter shading corresponds to more localized states.}
\end{figure}
lattice potential and its magnitude can be roughly estimated by
applying Gaussian approximation for the Wannier states:
\begin{equation}
V\approx \frac{V_1}{2}{\rm exp}(-\frac{\alpha^2}{\sqrt{V_0/E_r}}).
\label{eq:veq}
\end{equation}
We note that $V$ depends on $V_1$, $\alpha$, and $V_0/E_r$.
As a na\"ive extension to the AA model, we ask what will happen if
the coupling between next-nearest-neighbors is included. To
answer this question, we consider the model:
\begin{equation}
\sum_{d=1,2}t_d(u_{n-d}+u_{n+d})+V_nu_n=Eu_n \label{eq:t1t2model}
\end{equation}
where $V_n = V\cos(2\pi n)$. We solve the equation by direct
diagonalization. To quantify the localization of the wave
function, we compute the inverse participation ratio (IPR):
\begin{equation}
IPR^{(i)}=\frac{\sum_n|u^{(i)}_n|^4}{({\sum_n|u^{(i)}_n|^2})^2},
\label{eq:ipr}
\end{equation}
where the superscript $i$ denote the $i$-th eigenstate (ordered
according to energy from low to high). For spatially extended
states, IPR approaches zero whereas it is finite for localized
states \cite{Kramer93}.
\begin{figure}
\includegraphics[width=.45\textwidth]{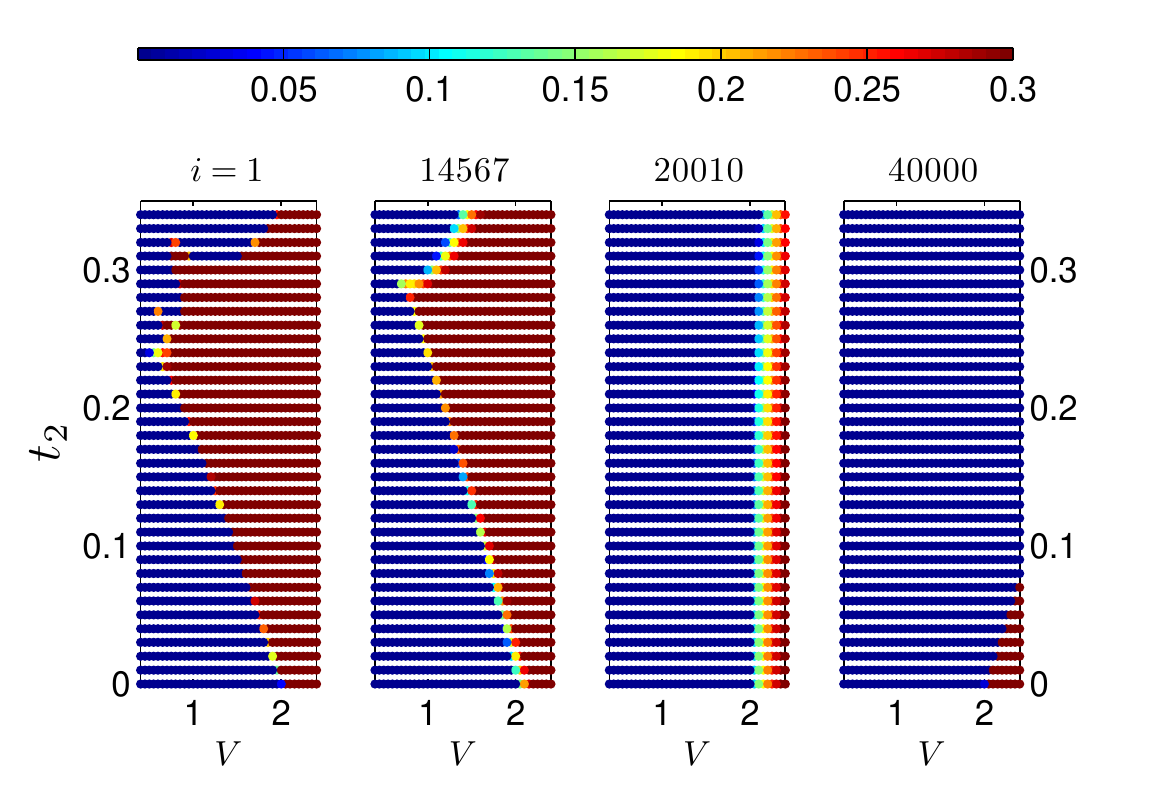}

\caption{\label{fig:Fig2} Inverse participation ratio on the
$t_2-V$ plane for $\alpha =(\sqrt{5}-1)/2$ based on the $t_1-t_2$
model. The four panels correspond to four eigenstates labeled by
$i$, with ascending eigenenergies. Blue regions correspond to more extended states and red regions correspond to more localized states.}
\end{figure}

Fig. \ref{fig:Fig1} shows the IPR values of all eigenstates as a
function of the effective strength $V$ of the secondary lattice
based on the tight-binding $t_1-t_2$ model with $\alpha =
(\sqrt{5} - 1)/2$ for various values of $t_2$ ($t_1$ is chosen to be unit of energy). The calculation for
Fig. \ref{fig:Fig1} is done for a system with 1000 sites in the
primary lattice. For small values of $t_2$ (e.g. $t_2$ = 0.01),
the localization properties of the system have essentially the
same features as those determined by the AA model. However, when
$t_2$ = 0.05 or higher, AA duality is clearly destroyed and localization transitions appear to be energy dependent.  For lower energies, the transition can appear for $V<2t_1$ and for higher energies, the transition can appear for $V>2t_1$.

In order to demonstrate the dependence of the localization transition on $t_2$ , we show the distribution of IPR on the
$t_2-V$ plane for four different eigenfunctions with $\alpha =
(\sqrt{5} - 1)/2$ in Fig. \ref{fig:Fig2}. For the
calculation, the size of the system is chosen to be 40,000. At
$t_2$ = 0, the $t_1-t_2$ model reduces to the AA model, and
from Fig. \ref{fig:Fig2}, one can see the sharp transition when $V$
is increased across the duality point $V = 2$. However, the
localization property of the system is greatly complicated when
$t_2$ is finite. Besides the appearance of mobility edges, our results also reveal that the dependence of the
localization property on $t_2$ is not monotonic, e.g. at fixed $V
< 2$ when $t_2$ is increased the ground state could be tuned from
extended to localized, but further increasing of $t_2$ could bring
the ground state into an extended state again.

We infer from the results presented in Figs. \ref{fig:Fig1} and \ref{fig:Fig2} that 1) the AA duality is destroyed by having $t_2 \ne 0$; 2) instead of the $V=2t_1$ dual point, the system has energy dependent mobility edges for $t_2 \ne 0$; 3) the precise localization condition deviates up or down from the $V=2t_1$ AA condition depending on the energy of the eigenstate and the value of $t_2$. As illustrated by Figs. \ref{fig:Fig1} and \ref{fig:Fig2}, the $t_1-t_2$ model itself could be of interest.
However, for the
study of localization properties in 1D incommensurate lattices,
its validity must be dealt with caution, especially when $t_2$ is
not sufficiently small as compared with $t_1$. The tight binding
nn and nnn hoping integrals $t_1$ and $t_2$ can be estimated with
the Wannier basis, which is fully determined by the primary
lattice. One can easily estimate that when $V_0 = 3E_r$, the ratio of $t_2/t_1$ is on
the order of $10\%$. To get higher $t_2/t_1$ ratio, one will need
to tune the lattice potential shallower and should expect the
tight-binding approximation to break down at some
point. Alternatively, to study the interesting physics of
localization in this regime, we numerically solve the
single-particle Schr\"{o}dinger equation without any tight-binding approximation:
\begin{equation}
(-\frac{\hbar^2}{2m}\frac{d^2}{dx^2}+V(x))\psi(x)=E\psi(x).
\label{eq:schrod}
\end{equation}
To achieve this goal, we discretize the Schrodinger equation in
the position basis with a finite system size of length $L = Na$,
where $a$ is the lattice constant of the primary lattice
associated with $V_0$. The continuous Schr\"{o}dinger equation is
now cast into the following form:
\begin{eqnarray}
(-\frac{\hbar^2}{2m})\frac{\psi_{n+1}-2\psi_n+\psi_{n-1}}{\delta^2}+\nonumber\\
(V_0\cos(2k_Ln\delta)+V_1\cos(2k_L\alpha n\delta))\psi_n=E\psi_n,
\label{eq:dschrod}
\end{eqnarray}
where $\delta = Na/M$ is the step interval for the discretization
with $M$ denoting the total number of steps. Then we proceed by
diagonalizing the $M\times M$ matrix of the discretized
Hamiltonian and study the first $N$ eigenstates with smallest
energy eigenvalues. These states would correspond to the ground
band for the case with no secondary lattice (i.e. $V_1 = 0$). In
our calculations for the following results, we have set $N = 500$,
$M = 80,000$, and $2k_L$= 1.

\begin{figure}
\includegraphics[width=.5\textwidth]{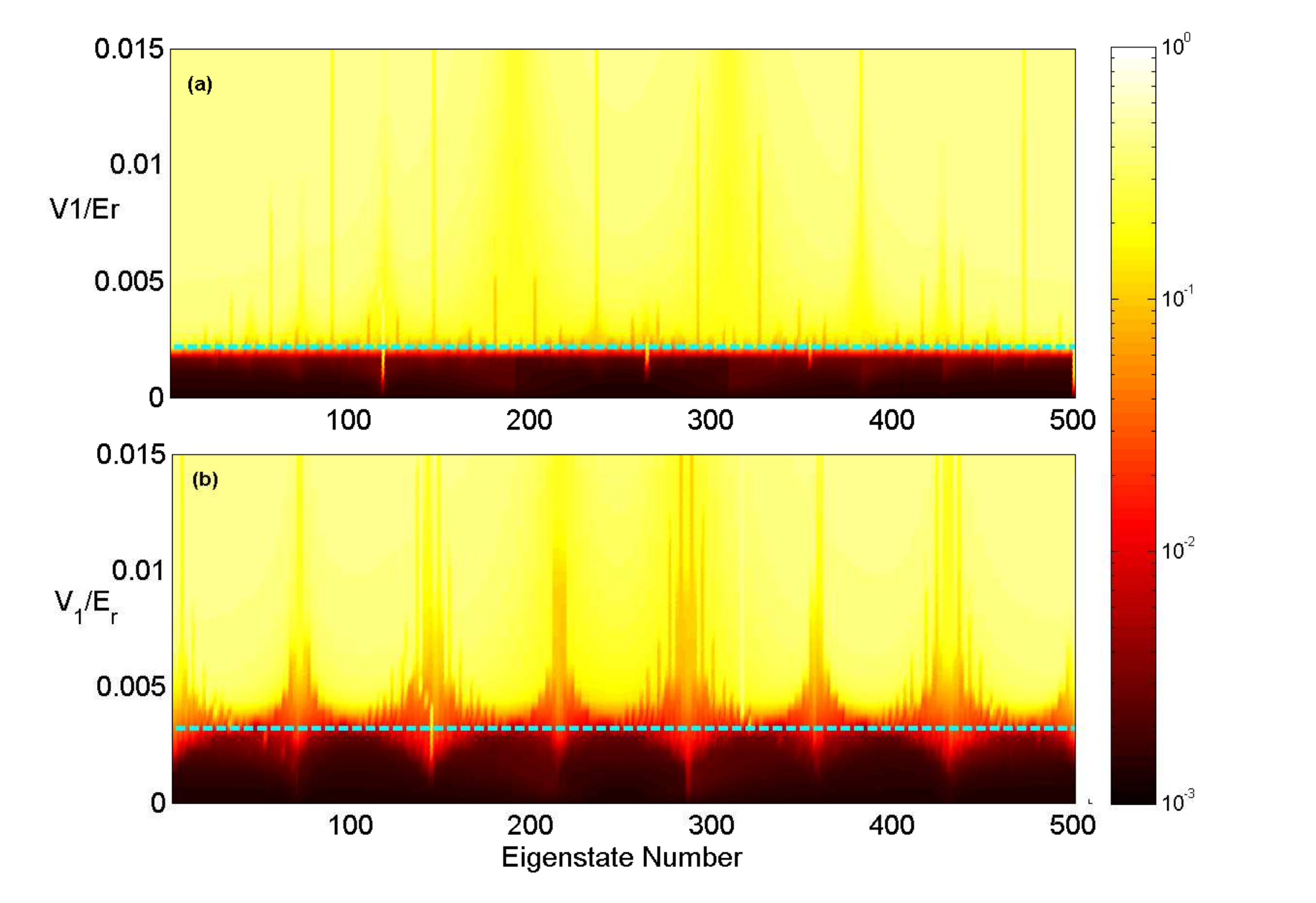}

\caption{\label{fig:Fig3} Inverse participation ratio obtained by
solving the Schr\"odinger equation and calculated AA duality
point (dashed line) at $V_0 = 30E_r$ (a) $\alpha=(\sqrt{5}-1)/2$;
(b) $\alpha=\pi/2$.}
\end{figure}

IPR values (obtained with Eq.( \ref{eq:ipr}) by replacing $u_n$
with $\psi_n$) of the first $N$ eigenstates as a function of the
secondary lattice strength $V_1$ are shown in Fig. \ref{fig:Fig3}
for a primary lattice strength of $V_0 = 30E_r$. In Fig.
\ref{fig:Fig3}(a) the irrational ratio $\alpha$ is set to be the
inverse golden mean, $(\sqrt{5} - 1)/2$ whereas in Fig.
\ref{fig:Fig3}(b), $\alpha = \pi/2$. The bold-dashed line
represents the AA duality point calculated with Eqs.
(\ref{eq:teq}) and (\ref{eq:veq}). We can see that the
localization properties shown in Fig. \ref{fig:Fig3} closely
resemble the well-known results from the AA model (see top panel in Fig. \ref{fig:Fig1}).  We do note, however, that the IPR results of Fig. \ref{fig:Fig3} indicate a dependence on the specific value of $\alpha$ with $\alpha=(\sqrt{5}-1)/2$ providing a sharper AA duality than $\alpha = \pi/2$.

\begin{figure}
\includegraphics[width=.5\textwidth]{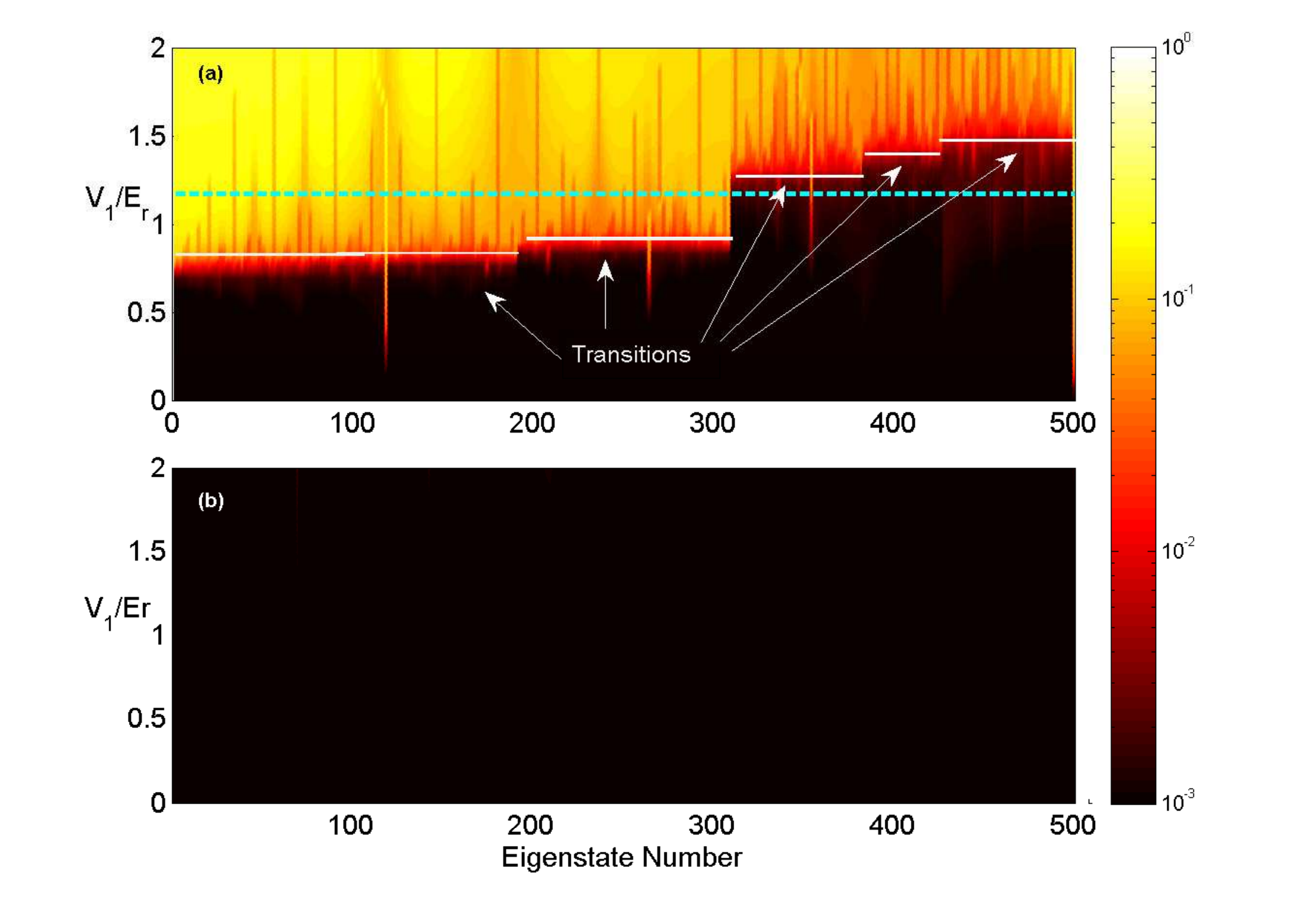}

\caption{\label{fig:Fig4} (a) Inverse participation ratio obtained
by solving the Schr\"odinger equation and calculated AA duality
point (dashed line) at $V_0 = 2E_r$ (a) $\alpha=(\sqrt{5}-1)/2$,
solid lines are estimated location of localization transitions;
(b) $\alpha=\pi/2$.}

\end{figure}
\begin{figure}
\includegraphics[width=.5\textwidth]{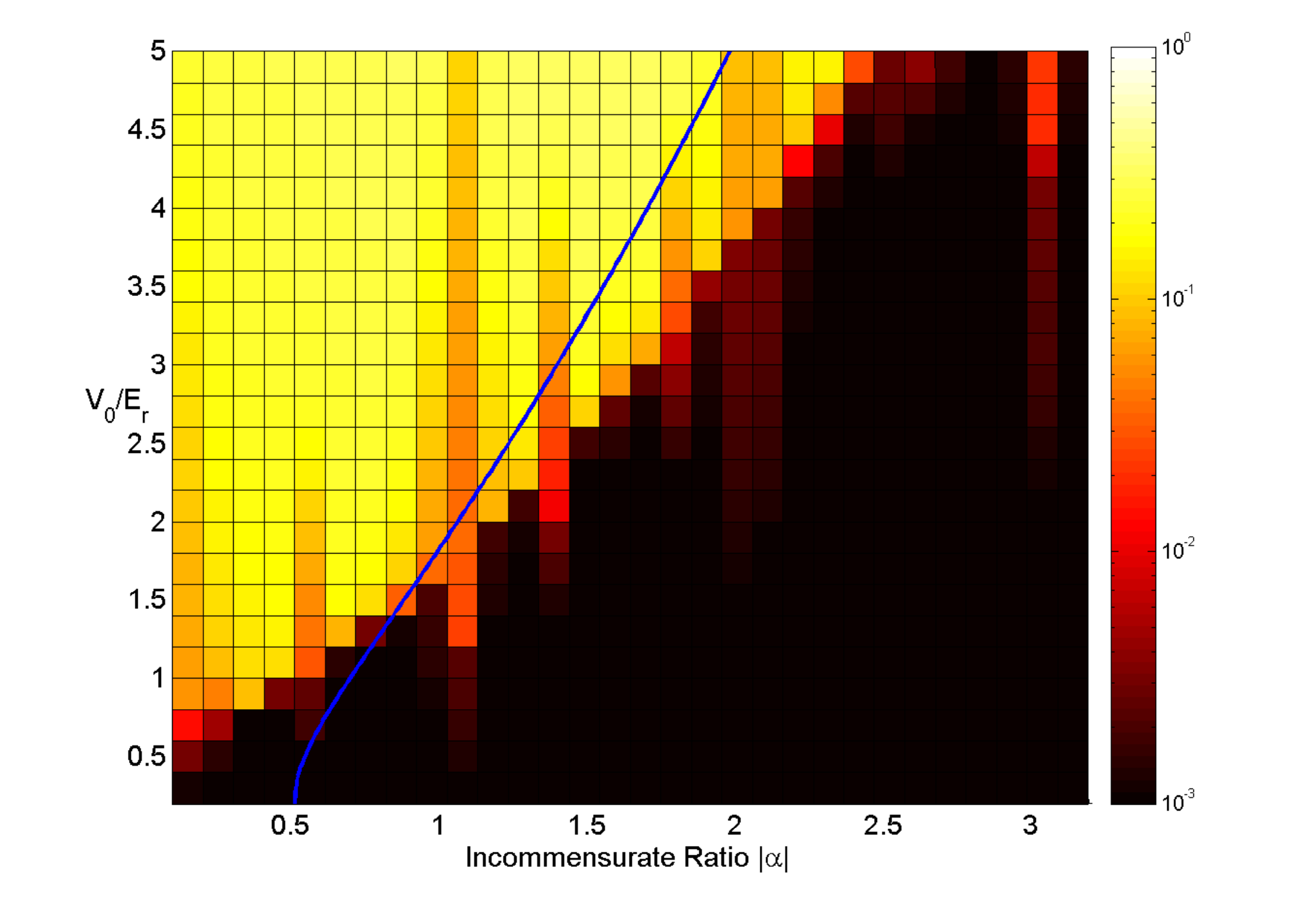}

\caption{\label{fig:Fig5} Inverse participation ratios of the ground
state wavefunction for the case $V_0 = V_1$ and $\alpha$ equal to fractional
multiples of $(\sqrt{5}-1)/2$.  The solid blue line represents an approximate analytical boundary between localized and extended regions based on the AA duality point.}

\end{figure}
In Fig. \ref{fig:Fig4}(a) we show the IPR values for the case of
$V_0 = 2E_r$ and $\alpha = (\sqrt{5} - 1)/2$. In this case, the
eigenstates no longer appear to localize all at once, but in
discrete steps (represented by the solid lines in the figure).
This localization behavior is similar to what we observed in
the $t_1-t_2$ model (see bottom panel in Fig. \ref{fig:Fig1}). Also the transitions
occur at fairly large values for $V_1$, where the secondary
lattice can no longer be treated as a perturbation. We have also
studied the cases where $V_0 = 2E_r$, $\alpha = \pi/2$ (Fig. \ref{fig:Fig4}(b)) and $\alpha
= (\sqrt{5}+1)/2$ (not shown in the figure). In these cases no
localization was observed in the eigenfunctions for any value of
$V_1$ investigated (up to $V_1 = V_0$). This suggests that
incommensurability between the lattices is not a sufficient
condition to observe localization for shallow cases.
 
To examine the dependence of the localization transitions on $\alpha$, we set
$V_0=V_1$ and calculate the IPR of the ground state for various
values of $V_0$ and $\alpha$ (the values of $\alpha$ examined are
all proportional to $(\sqrt{5} - 1)/2)$. These results are
shown in Fig. \ref{fig:Fig5}. We see fairly
distinct regions of localized and extended states, with
localization tending towards areas of larger values for $V_0$ and
smaller magnitudes for $\alpha$. The blue line in Fig. \ref{fig:Fig5} represents the set of points ($\alpha$,$V_0$) such that the AA duality point (calculated from Eqs. (\ref{eq:teq}) and (\ref{eq:veq})) is equal to the lattice strength $V_0$.  These sets of points serve as a simple heuristic estimation of the boundary between localized and extended states based on AA duality condition. 
Although in principle we should not expect the AA duality point obtained from Eqs.
(\ref{eq:teq}) and (\ref{eq:veq}) to be applicable in the case of shallow lattices, this simple analytical result is in good qualitative agreement with our numerical findings.

\begin{figure}
\includegraphics[width=.45\textwidth]{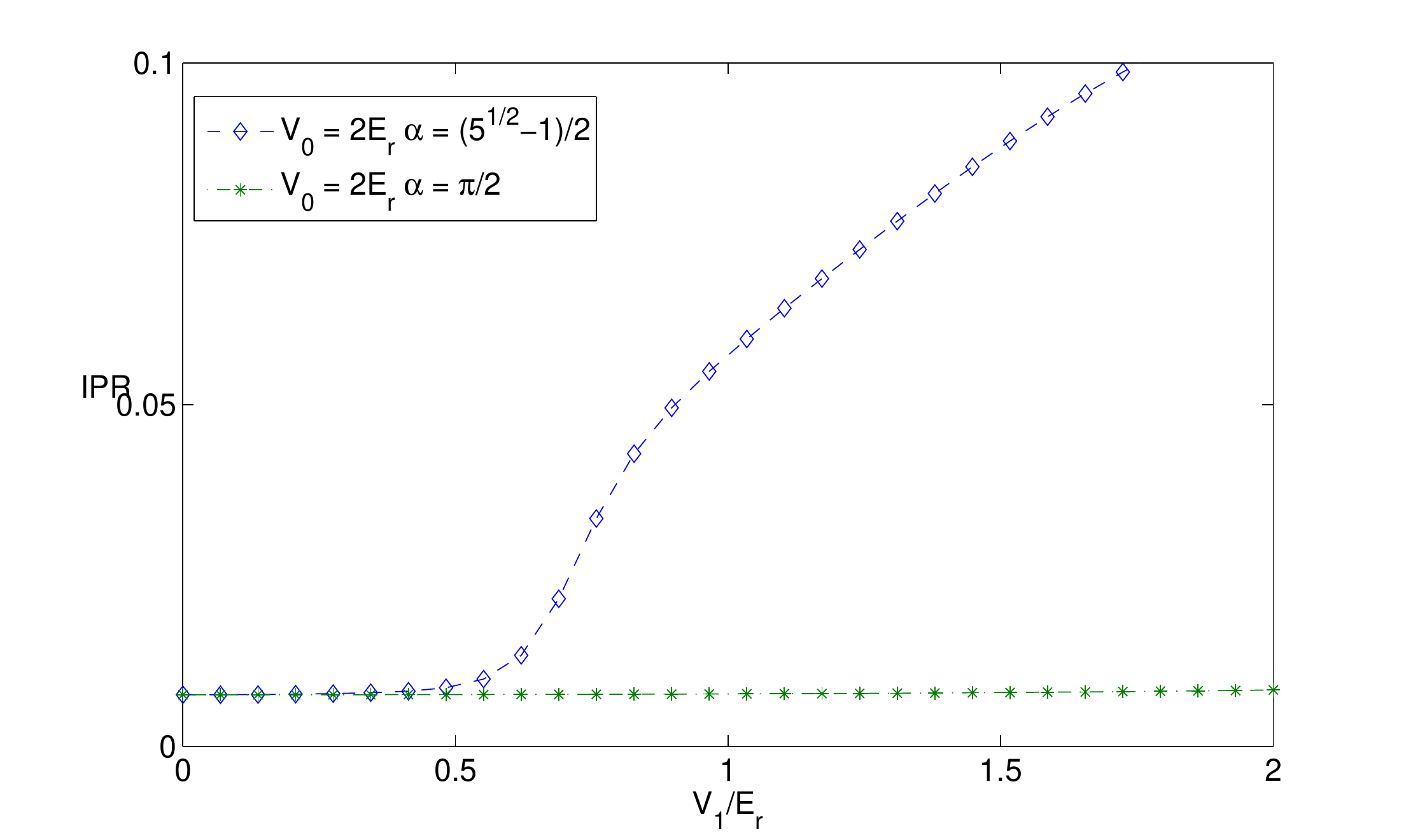}

\caption{\label{fig:Fig6} Inverse participation ratio of ground
state wavefunction at time $T_0\approx \hbar/E_r$ after the trap
potential $V_{\mathrm {trap}}=\Omega x^2$ has been turned off
($\Omega/E_r\approx 10^{-7}$).}

\end{figure}

We now briefly discuss how some of these results may be observed
in cold atom experiments. We consider a diffuse BEC that is loaded into an
incommensurate optical lattice, confined by a harmonic trap,
$V_{trap}=\Omega x^2$. We assume that the diffuse gas is prepared in
the ground state.  At time $T = 0$, the harmonic
trap is suddenly turned off and the BEC is allowed to diffuse.
Localization can be observed by monitoring the IPR of the density
wave function over time. In Fig. \ref{fig:Fig6}, we present the
calculated values for the IPR as a function of $V_1$ for the wave
function after a fixed period of time, $T_0\approx \hbar/E_r$, has
passed since the trap was turned off for the cases with $V_0 =
2E_r$, $\Omega/E_r\approx 10^{-7}$, $\alpha=(\sqrt{5}-1)/2$ and
$\alpha=\pi/2$. In the figure, we see the two cases are similarly
delocalized for small values of $V_1$. But for larger values of
$V_1$, the IPR for the $\alpha=(\sqrt{5}-1)/2$ case begins to
grow, showing increasing degree of localization, while in the
$\alpha=\pi/2$ case it remains constant.

In conclusion, we have studied the localization properties of
noninteracting particles in a one-dimensional incommensurate
optical lattice system based on a tight-binding $t_1-t_2$ model
with nearest-neighbor as well as next-nearest-neighbor hopping. We
reveal the emergence of mobility edges when the
next-nearest-neighbor hopping is finite. We have also gone beyond
the tight-binding approximation by directly modeling the system
with the fundamental single-particle Schr\"odinger equation, which
is expected to provide more reliable theoretical description of
the system especially for the case with shallow primary lattice
potential. By diagonalizing the discretized Hamiltonian, we
numerically solve the Schr\"odinger equation. Our results clearly
show the existence of mobility edges. Our study also reveals that
the emergence of localization is sensitive to the magnitude of
the irrational ratio $\alpha$ of the incommensurate lattice
potentials when the system is well outside of the tight-binding
regime.  Our results also establish the fragile nature of the AA duality which gives way to mobility edges as soon as longer range hopping, even at the nnn level, is turned on.  It will be interesting to verify our predictions about the sensitive qualitative dependence of 1D incommensurate localization on $V_0$, $V_1$, $E_r$, and $\alpha$ through experiments in cold atomic systems \cite{Billy08, Edwards08, Roati08}.

\begin{acknowledgments}
This work is supported by ARO-DARPA-OLE and NSF-JQI-PFC.

\end{acknowledgments}


\end{document}